\documentclass[sigconf]{acmart}

\usepackage{balance}

\def\BibTeX{{\rm B\kern-.05em{\sc i\kern-.025em b}\kern-.08emT\kern-.1667em\lower.7ex\hbox{E}\kern-.125emX}}

\usepackage{balance}
\usepackage[utf8]{inputenc}
\usepackage{xcolor}
\usepackage{etoolbox}
\AtBeginEnvironment{quote}{\itshape}
\usepackage{amsthm}

\usepackage{hyperref}

\newcommand{\LE}{LifeExp~}
\newcommand{\DP}{Dep~}
\newcommand{\LF}{LifeFavor~}
\newcommand{\DF}{DepFavor~}
\newcommand{\LFQ}{LifeFavorQ~}
\newcommand{\DFQ}{DepFavorQ~}
\newcommand{\LFAI}{LifeFavorAI~}
\newcommand{\DFAI}{DepFavorAI~}
\newcommand{\LFPsy}{LifeFavorPsy~}
\newcommand{\DFPsy}{DepFavorPsy~}

\AtBeginEnvironment{quote}{\itshape}

\begin{document}

\fancyhead{}

\title{Artificial Artificial Intelligence: Measuring Influence of AI `Assessments' on Moral Decision-Making}


\author{Lok Chan}
\affiliation{\institution{Duke University}}
\email{lok.c@duke.edu}

\author{Kenzie Doyle}
\affiliation{\institution{Duke University}}
\email{kmd87@duke.edu}

\author{Duncan C. McElfresh}
\affiliation{\institution{University of Maryland}}
\email{dmcelfre@math.umd.edu}

\author{Vincent Conitzer}
\affiliation{\institution{Duke University}}
\email{conitzer@cs.duke.edu}

\author{John P. Dickerson}
\affiliation{\institution{University of Maryland}}
\email{john@cs.umd.edu}

\author{Jana Schaich Borg}
\affiliation{\institution{Duke University}}
\email{js524@duke.edu}

\author{Walter Sinnott-Armstrong}
\affiliation{\institution{Duke University}}
\email{walter.sinnott-armstrong@duke.edu}

%
\renewcommand{\shortauthors}{Chan, et al.}

%
\begin{abstract}
Given AI's growing role in modeling and improving decision-making, how and when to present users with feedback is an urgent topic to address. We empirically examined the effect of feedback from false AI on moral decision-making about donor kidney allocation. We found some evidence that judgments about whether a patient should receive a kidney can be influenced by feedback about participants' own decision-making perceived to be given by AI, even if the feedback is entirely random. We also discovered different effects between assessments presented as being from human experts and assessments presented as being from AI.
\end{abstract}



%


%
\maketitle

\section{Introduction}

As AI becomes more prevalent as a tool to evaluate past choices and improve future decision-making, how and when to present users with feedback will become an increasingly urgent topic to address. Determining user susceptibility to inaccurate feedback from AI will also be crucial to prevent baseless---or malevolent---intervention in decision-making assistance. In this work we aim to elucidate how people respond to feedback given by AI about their own decision-making, to identify opinions about the credibility of AI feedback, and to measure susceptibility to unfounded AI assessments.

We investigate how the \emph{perception} of an AI-generated assessment impacts moral decision-making. Specifically, we consider the effect of AI \emph{assessments}: people's responses to \emph{predictions} about what they \emph{will} do, from \emph{assessments} of the kind of person \emph{they are} \cite{oyserman2011}. From a modeling perspective, this distinction is not as pronounced---agents are defined by their actions and desires. However, previous research has shown that people's motivation for future action is often shaped by their perception of who they are \cite{oyserman2009}.

Thus, as a starting point for a psychological investigation, we focus on the effect an output has when interpreted as \emph{an assessment} of one's moral values. In addition to our primary endeavor, we consider two sub-questions: Do people respond more to assessments that they believe to have come from AI than to those that they believe to have come from a human? What is the effect of this perception if they are immediately asked to report if they agree or disagree with the assessment? 

To address these questions, we conducted three studies on the effect of what we call an \emph{artificial artificial intelligence (AAI)} assessment, in which random statements about users' values were (falsely) presented as AI-generated feedback. In each study, participants received an AAI assessment of their morality before they were presented with a series of moral dilemmas involving kidney allocation. We found that AAI assessments had an effect on participants' allocation choices between patients. Under some conditions, this effect was slightly altered if participants were first asked whether they \emph{agreed} with the assessment. We also found differences between the effect on people who believed the assessment to be AI-generated, compared to those who believed that it was from human experts.

As our studies build on research across several disciplines including computer science, moral psychology, and social psychology, part of our effort is to suggest how relevant concepts can be reasonably translated across these frameworks. From the perspective of computer science, our question about AI assessment and moral decision-making can be interpreted as determining the effect of intervening on preferences by introducing a prediction, which is presented to the user as an assessment in the sense described above. The question of how most effectively to learn preferences is the focus of \emph{preference elicitation}, a field with broad applications in several fields including medicine~\cite{weernink2014systematic}, marketing~\cite{huber1993effectiveness}, and auction design~\cite{conen2001preference}. Preference elicitation is also a primary concern in \emph{social choice}~\cite{arrow2012social}---the study of how to aggregate preferences for collective decision-making; social choice has received significant attention from computer scientists~\cite{brandt2016handbook}, raising further questions about automation and AI in human decision-making.

However, to our knowledge, the effect of the \emph{perception} of these predictions has not been covered in the computer science or preference elicitation literature. To study this, our `artificial AI' (AAI) goes one step further: after making an assessment of a user, it presents this assessment \emph{directly to the user}. Does this impact the user's moral decision making, interpreted as an expressed preference? This is an area in which research from psychology could be informative.

When people receive assessments of themselves, even when the assessments have no basis in reality, their behaviors may still be influenced. For instance, behavior sometimes changes in response to and accordance with false feedback: random feedback about results of a political implicit associations test can impact level of political engagement \cite{vitriol2019}, and the Barnum (or Forer) effect, wherein people accept vague assessments of their personality as true regardless of accuracy, is well documented \cite{furnham1987}. Furthermore, people can be prompted to revise their political opinions when presented with an inaccurate summary of relevant decisions that they \emph{just} made \cite{hall2013}. The effect of false reporting of individuals’ own beliefs is especially enduring if they provide an explanation for the position they were informed they took initially, despite that being the opposite of their original position  \cite{strandberg2018}. The present research addresses whether people's beliefs about their own opinions is susceptible to false influence from AI.

\section{Methodology}\label{sec:methods}

We designed a custom online platform to study the effect of AAI assessment on decision-making. Each decision took the form of a \emph{pairwise comparison}~\cite{bradley1952rank}---a decision format used widely in many disciplines, where an agent selects their most-preferred item from two options. Using a participant's answers to these comparisons, we learned their \emph{decision function} to describe their choices. 
In what follows, we briefly explain our methodology by describing the basic scenario we presented to participants, the decision function used for analysis, and the custom data-gathering platform that simulates the decision making environment. 

\paragraph{Scenario: Moral Decisions on Kidney Allocation}\label{sec:study-design} In order to elicit preferences efficiently---a decision function---from our participants, we designed a simplified choice scenario with a small class of easily-measurable preferences, situated within a hypothetical environment resembling a real life-and-death decision made every day: allocating donor kidneys. Many of the features that the general population considers important in determining kidney allocation go beyond objective medical facts and enter into ethical opinions. In the US, these decisions are guided strictly by UNOS policy\footnote{\texttt{\url{https://optn.transplant.hrsa.gov/media/1200/optn_policies.pdf}}}. However, the general population may think other features to be relevant to these decisions. For example, people are often unwilling to allocate organs to patients with features that do not contribute to organ failure or prognosis \cite{ubel1999}. This makes kidney allocation a valuable avenue for comparing attitudes toward different ethically relevant characteristics, and for studying differences between informal attitudes and lay opinions about what should be included in formal policies.

We selected life expectancy (henceforth, ``LifeExp'') and number of dependents ~(henceforth, ``Dep'') as the basis for comparison between patients in need of a kidney, because both have been demonstrated to be of importance to the general population for ethically relevant 
reasons \cite{freedman2018adapting} that can be manipulated independently of one another (i.e. the patient's life expectancy implies nothing about their number of dependents and vice-versa).  
We also included patient age, which is $40$ years for all patients. In these studies we presented each participant with several pairwise comparisons, where the alternatives are two patients in need of a single kidney; this is akin to the problem of allocating a single deceased-donor kidney to one of two patients. In each study, these features were explained to participants as follows:

\begin{description}
    \item[Life Expectancy] How many years the patient is expected to live if they receive the kidney transplant, if the patient makes no lifestyle changes. 
    \item[Dependents] The number of children under the age of 18 for whom the patient is responsible for providing at least half the necessary support, including food, shelter, and clothing.
    \item[Age] The current age of the patient. All patients in the scenarios are 40 years old. This feature does not vary.
\end{description}

Because \LE and \DP impose different types of value on a transplantable kidney, those two features may be varied independently without either implying anything about the other. We held age constant at 40 years to limit further the assumptions participants could make about the patients from the target features.

\paragraph{Measuring Participant Decision Functions: Feature Dominance} Each participant may have arbitrarily complicated preferences in this setting. One participant may only allocate kidneys to patients with \LE greater than 10 years, and choose randomly otherwise. Another participant might only care about a different feature (such as the patient's age) and completely ignore both \LE and Dep.  To avoid this problem we constructed a set of pairwise comparisons that essentially ask which \emph{feature} the participant cares most about.
In each comparison, one patient \emph{always} had greater \LE and less \DP than the other patient.

We assume that participants answer each comparison by selecting the patient with either greater \LE or Dep. Formally, we assume that each participant has a simple \emph{decision model}: with probability $p$ they prefer the patient with greater life expectancy, and with probability $1 - p$ they prefer the patient with more dependents. For ease of exposition, we express these probabilities as percentages, where
$$\text{\%Life}\equiv 100\times p$$ 
Because participants only considered a patient who was always strictly greater in life expectancy and fewer in dependents to another, this is simply \emph{the percentage of comparisons where the participant selects the patient with greater \LE}:

$$\text{\%Life}\equiv 100 \times \frac{\text{\# of \LE Favoring Decisions}}{\text{\# of Total Decisions}}$$


To measure the impact of an AAI assessment on participant decision functions, we first learned \%Life for each participant. We then compared the effect AAI assessments had on each intervention group by aggregating the participants' \%Life from respective groups.

\paragraph{Custom Online Platform for Data Collection} In a style similar to The Moral Machine Project,\footnote{\texttt{\url{http://moralmachine.mit.edu}}} we created a custom online platform to facilitate data collection, called Who Gets the Kidney?\footnote{\texttt{\url{https://whogetsthekidney.com}}}  The core component of the online platform is the sequential display of a set of a hypothetical decision-making scenarios in which participants choose one of two patients to receive a donor kidney. In each study, every participant received the same set of scenarios, but the display of each scenario was randomized: each scenario appeared in a different order for each participant, and each hypothetical recipient in a scenario was randomly selected to be presented on either the left or the right side. Participants were given a chance to review their answer, and the option to change their minds as needed.  


For each study below, participants were first given a brief description of the decision-making scenario (kidney allocation). They were informed that if one patient received the kidney, the other would \emph{not} receive one, and that if a patient did not receive the kidney they were expected to live less than a year. Furthermore, it was made explicit that all transplants were likely to be successful.

From pilot testing, we expected that participants recruited online would maintain attention on the kidney allocation task for between 20 and 30 patient profile pairs. As such, all studies asked participants to respond to 20 pairs to ensure their focus. Further, pilot testing suggested that decision-making time decreased substantially after the first three pairs, indicating that participants took about three decisions to get a grasp on the task. Therefore, all studies used at least 10 pairs to ensure that most of the participants’ decisions were made with complete familiarity with the task.



\section{Study 1}

\subsection{Method}


114 participants were recruited on Amazon Mechanical Turk (MTurk) in a single cohort (on a Monday afternoon). Only United States residents were used. After data collection was completed, 17 participants were excluded from analyses for failing an attention check which required them to report the assessment they received. 6 attempts were removed due to participants with the same IP address making multiple attempts. This leaves a final sample of N = 91 (41\% females and 59\% males; mean age = 37.7, SD = 11.2, 76\% white).


Participants were presented with background information on kidney allocation and about the patient features in this survey. On our online platform, participants were asked to make decisions on a set of 10 scenarios. To limit decision complexity, we further simplified the scenarios by keeping \emph{all but one patient feature} the same for each comparison. One patient always had a life expectancy of 20 years and 0 dependents, while the other patient had 4 dependents. The only variable feature was the life expectancy of the latter patient, ranging from 1 to 19 years.
After these 10 decisions, an ``assessment'' screen was displayed with the intervention text. Participants were randomly assigned to view one of the following AAI assessments:

\begin{description}
     \item[\LF]``According to our AI model, you care a lot about the life expectancy of the patients when making decisions about who will get a kidney.''
     \item[\DF]``According to our AI model, you care a lot about how many dependents patients have when making decisions about who will get a kidney.''
     
\end{description}
 Immediately afterwards, participants were prompted to make 10 more decisions. While the comparisons were the same as the initial 10, participants were not explicitly informed of this. Furthermore, the sequence in which the queries were shown was shuffled, and the sides on which patient profiles in each comparison were displayed were randomly switched.  After completing all kidney patient allocation choices, participants responded to a survey, which included demographic information and a question on whether they agreed with the AAI’s assessment. They were then debriefed, which included telling them that the feedback was actually random. 

\subsection{Results}

\begin{figure}
\centering

\includegraphics[width=8cm]{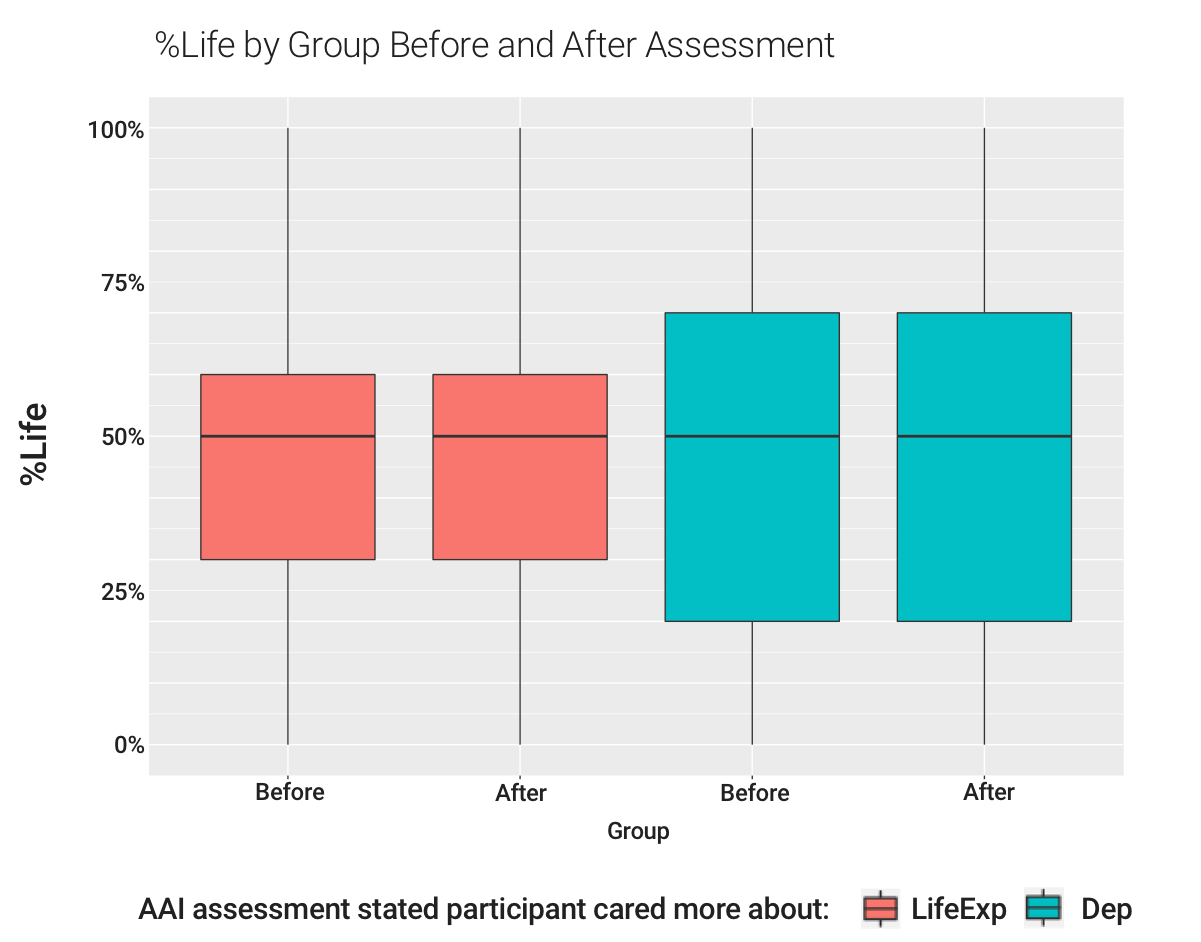}
\caption{Study 1: medians and first/third quartiles for \%Life, before and after assessment for each participant}
\label{study1b}
\end{figure}

After exclusions, 32 participants received the LifeExp intervention and 59 received the DepFavor intervention. \%Life was created as a summary variable to capture the proportion of life expectancy-favoring decisions for each participant. Figure \ref{study1b} displays the visualization of the result.

As the Shapiro-Wilk test showed that normality did not hold for \%Life ($p<0.01$), we used the non-parametric Wilcoxon rank sum test to determine if the \%Life for participants in each group moved toward their respective assessment after it was given. Specifically, we tested if, for \LE group, if the median \%Life for the identical set of comparison was \emph{higher} after assessment than before it. No evidence suggested that this was the case ($p=0.30$). The same was true for participants who received the \DF assessment: no evidence suggested that the median \%Life were lower, i.e., more dependents favoring, after they were told that they cared more about the number dependents ($p=0.54)$.

\subsection{Discussion} 

Strikingly, both pre-assessment and post-assessment groups are nearly identical. Two important observations came from this exploratory study. First, in the process of generating the initial 10 decisions as the input for the ``AI,'' participants may have formed explicit decision rules that made them resistant to intervention. This would have been simple to do, since one patient always had 20 years of life expectancy with no dependents in every comparison. Second, the lack of the effect of \DF compared to \LF could be attributed to the fact to that \DP was invariant across patients, while one patient's \LE varied between comparisons.


\section{Study 2}

In this study we modified the design of Study 1 to amplify potential effects of the intervention. First, we changed the position of the intervention relative to the allocation decisions. Choice tendencies that are constructed through repetition are more resistant to change than choice tendencies developed through contextual cues \cite{amir2008}. Presenting self-referential information before a task tends to cause behaviors consistent with that self-referential information \cite{oyserman2015}. To avoid pre-assessment heuristic development, and to promote the AAI assessment as salient self-relevant information, we presented the AAI intervention before the kidney allocation task. We asked participants to complete a task on which the AAI’s assessment could plausibly be based.
 
Second, we added two conditions in which we asked participants if they agreed with their assessment before and after the allocation task. In Study 1, participants could develop an opinion about their assessment while engaged in the allocation task, and report their end agreement or disagreement based on their experience. We predicted that judging the accuracy of the assessment before the task would amplify either incorporation of the assessment into participants' self-referential beliefs (if they agreed), or reactions in opposition to the assessment (if they disagreed).

\subsection{Method}



350 participants were recruited on MTurk in two cohorts (one around midday on a Monday, one during afternoon on a Wednesday), although they were randomly assigned among all groups within each cohort. As in Study 1, only US residents were used. 7 participants could not accurately report assessment received, and 11 failed an attention check implemented to exclude participants who did not engage with the task (at the risk of also eliminating participants who deliberately chose to allocate at random) by allocating a kidney to a less desirable patient based on both features  
(0 dependents, 1 year life expectancy, versus 0 dependents, 20 years life expectancy). 11 participants failed to finish the task. 6 attempted the experiments twice. Some participants belonged to more than one of the categories above. The final sample after exclusions was therefore N = 322 (41\% female and 58\% males; mean age = 35.8, SD = 10.2, 79\% white).

As in Study 1, participants first received background information about the task and patient features. They were also told that they would answer a series of questions that an AI agent would use to make an assessment about what they found most important in the kidney allocation task.

Participants rated, on a 5-point Likert-type scale, the extent to which they agreed or disagreed with 14 statements about the importance of using example features to determine who should receive a kidney (e.g., \emph{``I feel that race is important in determining which patient should receive a kidney''}). These example features \emph{did not} include life expectancy, number of dependents, or age. Afterwards, participants were randomly assigned to one of five conditions, in which all but control received AAI assessments that used contrastive language (e.g., \emph{``you care more about the life expectancy of the patients than how many dependents they have''}):

\begin{itemize}
    \item \textbf{Control:} no assessment ($n=65$)
    \item \textbf{\LF:} participants ``assessed'' by AAI to prioritize life expectancy over number of dependents ($n=66$)
    \item \textbf{\LFQ:} participants ``assessed'' by AAI to prioritize life expectancy over number of dependents, and asked immediately upon viewing the assessment if they agreed with it ($n=60$)
    \item \textbf{\DF:} participants ``assessed'' by AAI to prioritize number of dependents over life expectancy ($n=64$)
    \item \textbf{\DFQ:} participants ``assessed'' by AAI to prioritize number of dependents over life expectancy, and asked immediately upon viewing the assessment if they agreed with it ($n=67$)
\end{itemize}

Following intervention, all participants responded to twenty curated patient comparisons in random order. As before, each comparison had one patient with greater \DP and one with greater LifeExp; unlike Study 1, both \LE and \DP varied by patient and pair. Participants then completed the same post-task survey as in Study 1, and were debriefed as in Study 1.

\subsection{Results}

\begin{figure}
\centering
\includegraphics[width=8cm]{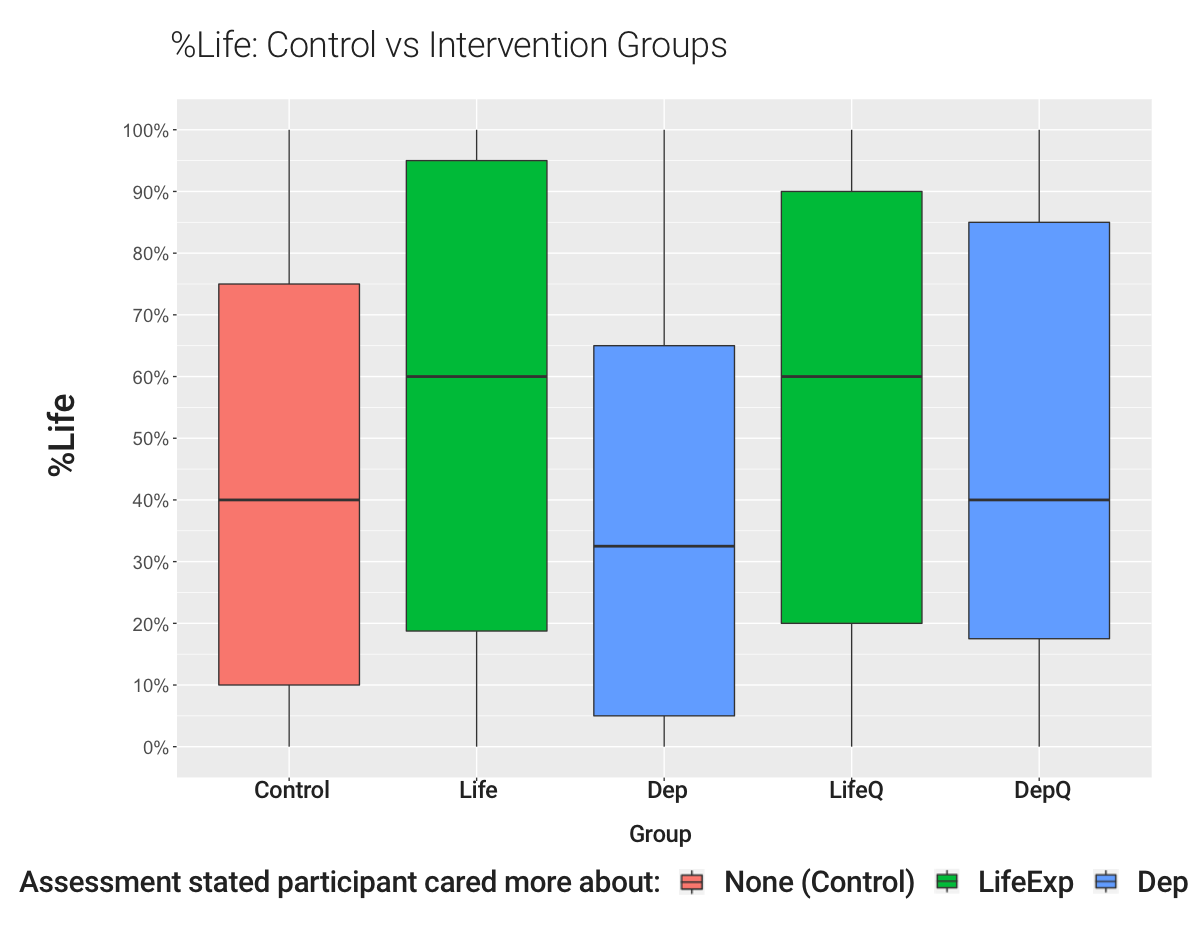}
\caption{Study 2: medians and first/third quartiles for \%Life of participants in each group}
\label{fig:study2:pd}
\end{figure}

As described in Methodology, we found \%Life for each participant by calculating the percentage of comparisons in which they decided to allocate the kidney to the patient with higher life expectancy. Figure~\ref{fig:study2:pd} shows a box plot of \%Life for each group, aggregated over all comparisons.

Intervention groups \LF (M=56, SD=10) and LifeFavorQ (M=56, SD=10) both had greater mean \%Life than Control (M=45, SD=13). Also consistent with our main hypothesis, \DF (M=40, SD=10) had lower mean \%Life than Control. These results are suggestive of the hypothesis that an AAI assessment influences people to make decisions aligned with the assessment. The exception, however, is that \DFQ (M=49,SD=13) had a \emph{higher} \%Life than Control. 

Our outcome measure, \%Life, once again violated the normality assumption (Shapiro-Wilk test: $p<0.01$). As a result, we used the non-parametric Wilcoxon rank sum test for one-sided comparison between intervention groups and the control. Compared to the control group, we have some reasonable evidence suggesting that the median \%Life was higher for \LF($p=0.056$) and \LFQ ($p=0.057$). However, we saw no evidence suggesting that the median \%Life are lower for \DF($p=0.307$) and \DFQ ($p=0.85)$ than for Control.

\subsection{Discussion}

Overall, we may reasonably suggest that the modification of experimental conditions from Study 1 to Study 2  enhanced the effect of AAI assessments, as evidenced by the moderate directional results. In Study 1, the \LE assessment had virtually no impact on decision-making, but, in Study 2, participants in groups that received a \LF and \LFQ assessment favored patients somewhat more heavily on the basis of \LE  than control.  This could be due to participants receiving their assessments \emph{before} having the chance to develop decision preferences, but also could be due to the increased complexity of the comparisons. The result is not overwhelming, however: in comparison, there seems to be no evidence that \DF assessment had an impact on participants' decision-making. Even more curious is the high p-value($0.85$) from the comparison between \DFQ \%Life  and Control \%Life.

An exploratory analysis on the effect of participants’ agreement or disagreement with their AAI assessment prior to decision making could provide some insight: we partitioned the Q groups into 4 categories by assessment (LifeFavor/DepFavor) and response (Agree/Disagree). After calculating their respective Life\%, the DepDisagree group had the highest \%Life (M=81, SD=25, n=21), followed by ExpAgree (M=69, SD=30, N=49), then DepAgree(M=34; SD=30, n=46), and lastly ExpDisagree (M=19, SD=23, n=18). In other words, those who disagreed with their assessment seemed to make decisions contrary to their AAI assessment. However, we urge caution against drawing strong conclusions from these exploratory results.

\section{Study 3}

Study 3 compared the influence of AAI assessments to that of supposedly human assessments: we examined the effect of AAI assessments  relative to assessments believed to be generated by human experts.

\subsection{Method}

450 participants were recruited on MTurk in two cohorts (one between a Thursday evening and a Friday around midday, and one around midday on a Saturday), although they were randomly assigned among all groups within each cohort. As in Studies 1 and 2, only US residents were used. Exclusion procedures were similar to procedures in Study 2, except here we also checked if they could report the source of the assessment. 59 participants were excluded because they could not accurately report the assessment received or its source, 22 were excluded because of failed attention checks or failure to complete the task. The final sample after exclusions was therefore N = 369  (43\% female, 56\% male, and 1\% other/not indicated; mean age = 38, SD = 11.3, 73\% white). 

Methods and assessments were similar to in Study 2. Participants were randomly assigned to five groups:

\begin{itemize}
    \item \textbf{Control:} no assessment ($n=77$)
    \item \textbf{LifeFavorAI:} participants were given an AAI assessment stating that they care more about life expectancy than number of dependents ($n=80$)
    \item \textbf{DepFavorAI:} participants were given an AAI assessment stating that they care more about number of dependents than life expectancy ($n=75$)
    \item \textbf{LifeFavorPsy:} participants were informed that, based on a test made by ``expert psychologists,'' they care more about life expectancy than number of dependents ($n=74$)
    \item \textbf{DepFavorPsy:} participants were informed that, based on a test made by ``expert psychologists,'' they care more about the number of dependents than life expectancy ($n=63$)
\end{itemize} 

After the assessment, participants were told that their responses to comparisons would be used either to train an AI that models their decision-making, or by expert psychologists to develop a psychological test. The post-task survey was similar to the post-task survey used in Studies 1 and 2, as was the debriefing information.

\subsection{Results}

\begin{figure}
\centering
\includegraphics[width=8cm]{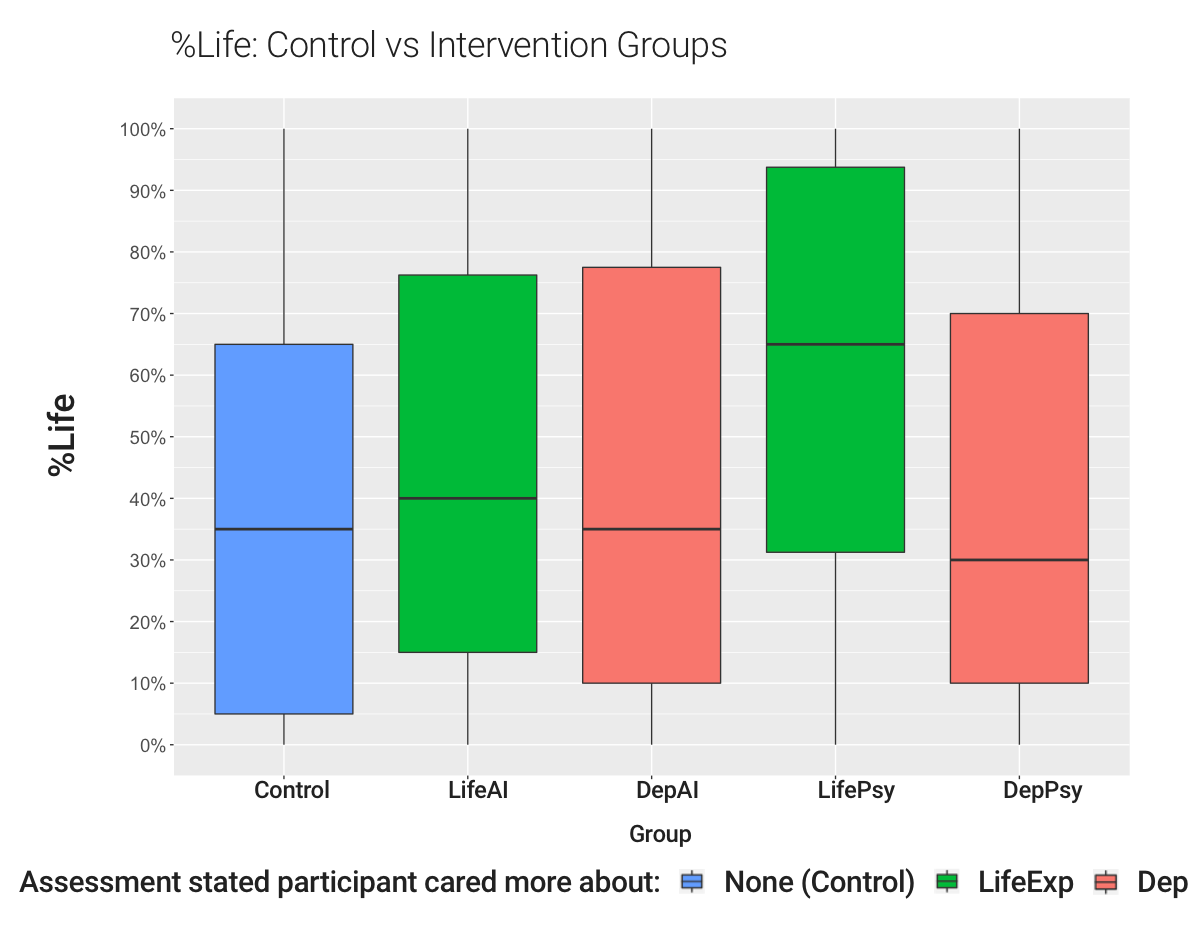}
\caption{Study 3: medians and first/third quartiles for \%Life for over all comparisons}
\label{fig:study3bp}
\end{figure}

As in Study 2, we calculated \%Life for each participant over all comparisons. Both groups that received LifeExp assessments---LifeFavorAI (M=45,SD=11) and \LFPsy (M=60, SD=11)---had higher \%Life than Control (M=40,SD=12). Inconsistent with Study 2, however, was that the \%Life were also higher for the groups that received dependent-favoring assessments---\DFAI(M=42, SD=11) and DepFavorPsy (M=42, SD=11). Figure~\ref{fig:study3bp} shows a boxplot of \%Life for subjects in each group.

Once again normality did not hold for \%Life (Shapiro-Wilk test: $p<0.01$). First, we started with two-sided tests to determine whether assessments from AAI had a difference in effect from assessments perceived to be from human experts. We found statistically significant evidence for the hypothesis that the median for \LFAI was not equal to \LFPsy ($p=0.01$), but there was no evidence of effect between \DFAI and \DFPsy ($p=0.93)$.

Next, we used the one-sided Wilcoxon rank sum test for comparison between intervention groups and the control. Consistent with the summary statistics, compared to Control \LFPsy had a statistically significant higher \%Life than Control ($p<0.001$). Next,  the hypothesis that \%Life for \LFAI was greater than Control with a low but statistically insignificant p-value ($0.15$). We found no statistical evidence in support of the hypothesis that subjects in \DFPsy($p=0.75$) and \DFAI($p=0.67$) allocated on the basis of dependents more, i.e., having lower \%Life, than control.

\subsection{Discussion}

We found strong evidence that, for the more impactful of our two patient features (LifeExp), there was a difference between the perception of ``AI'' and ``expert human'' assessments.

Although life expectancy-favoring AAI assessments influenced participants to make allocation decisions in the expected direction (as they did in Study 2) compared to control, participants responded more strongly to the ``expert'' life expectancy-favoring assessment by this metric than to any other condition. Furthermore, the results of the \LFAI and \LFPsy groups were significantly different from each other. This suggests that, for the more impactful of our two patient features, there was a difference between the perception of ``AI'' and ``expert human'' assessments.

Dependent-favoring assessments did not notably impact decision-making, although there was a very mild tendency for participants in the dependent-favoring assessment groups to make allocation decisions on the basis of life expectancy more frequently than control, which opposes our initial expectations. This could be related to the control participants in Study 3 making allocation decisions on the basis of dependents somewhat more frequently than did control participants in Study 2 (perhaps due to the hours or days at which data for Study 2 and Study 3 were collected). However, as the results were not strong in either direction in the dependent-favoring assessments, it is difficult to draw any firm conclusions about this intervention's effect.

\section{General Discussion}

Overall, we derive three central findings from these studies: First, life expectancy-favoring assessments had modest directional results on participants' decision-making. Second, dependent-favoring assessments had little notable effect. Third, the most statistically significant effect was that of the life expectancy favoring-assessment from ``expert psychologists,'' not that of any AAI assessment. 

That life expectancy-favoring assessments had some directional effect across all three studies, despite being artificial, suggests that AI assessments might be able to influence decision-making more with modifications to the life expectancy-favoring interventions. More accurate, genuine AI assessments might have a stronger effect, and a potential future direction would be to compare the effect of AAI assessments with those from a proper AI/ML model. However, it is also likely that the impact of assessments, whether from AI or AAI, could be intensified through other means as well, given that we were able to amplify the effects of the intervention between Study 1 and Studies 2 and 3 by changing the the order in which it appeared and the wording of the assessment. Methods of strengthening effects of assessments on behavior should be considered both in developing effective AI interventions and in exercising caution against the influence of inaccurate feedback. 

It is possible that participants were less receptive to the dependent-favoring assessment than the life expectancy-favoring assessment because it was a less complex variable and therefore easier to form an opinion about: each patient could have only up to four dependents, whereas they could have up to twenty years' life expectancy. Comparisons of different patient features could yield different results. 

The stronger influence of assessment from  ``expert psychologists'' compared to the AAI assessment in the life expectancy-favoring condition could be interpreted in several ways. As evidence-supported feedback is more effective in belief revision than unsupported feedback \cite{rich2017}, participants might have believed that evaluation about decision-making by a psychologist was better evidence of their preferences than an evaluation by AI. Alternatively, participants might have assigned more credibility to ``expert psychologists'' than to ``AI'' simply because of the word ``expert.''  

\paragraph{Acknowledgments} This work is supported by the project ``How to Use Artificial Intelligence to Enhance Human Moral Intelligence'' funded by the Templeton World Charity Foundation, ``How to Build Ethics into Robust Artificial Intelligence'' funded by Bass Connections, by NSF IIS-1846237, and by NSF IIS-1814056. We also thank members of the Moral AI Lab at Duke, including Claire Fu, Daniela Goya-Tocchetto, David Rein, and Gayan Seneviratna for feedback on this work.

%
\bibliographystyle{ACM-Reference-Format}
\balance
\bibliography{citation.bib}

%

\end{document}